\documentclass[aps,prl,preprint,groupedaddress,showpacs,superscriptaddress]{revtex4-1}

\usepackage{graphicx}
\usepackage{float,amsmath,verbatim,amssymb,stmaryrd}
\usepackage{epsfig}

\begin{document}


\title{Transverse electron-scale instability in relativistic shear flows}

\author{E. P. Alves}
\email{e.paulo.alves@ist.utl.pt}
\author{T. Grismayer }
\affiliation{GoLP/Instituto de Plasmas e Fus\~ao Nuclear, Instituto Superior T\'ecnico, Universidade de Lisboa, 1049-001 Lisbon, Portugal}
\author{R. A. Fonseca}
\affiliation{GoLP/Instituto de Plasmas e Fus\~ao Nuclear, Instituto Superior T\'ecnico, Universidade de Lisboa, 1049-001 Lisbon, Portugal}
\affiliation{DCTI/ISCTE - Instituto Universit\'{a}rio de Lisboa, 1649-026 Lisbon, Portugal}
\author{L. O. Silva}
\email{luis.silva@ist.utl.pt}
\affiliation{GoLP/Instituto de Plasmas e Fus\~ao Nuclear, Instituto Superior T\'ecnico, Universidade de Lisboa, 1049-001 Lisbon, Portugal}

\date{\today}

\begin{abstract}
Electron-scale surface waves are shown to be unstable in the transverse plane of a shear flow in an initially unmagnetized plasma, unlike in the (magneto)hydrodynamics case. It is found that these unstable modes have a higher growth rate than the closely related electron-scale Kelvin-Helmholtz instability in relativistic shears. Multidimensional particle-in-cell simulations verify the analytic results and further reveal the emergence of mushroom-like electron density structures in the nonlinear phase of the instability, similar to those observed in the Rayleigh Taylor instability despite the great disparity in scales and different underlying physics. Macroscopic ($\gg c/\omega_{pe}$) fields are shown to be generated by these microscopic shear instabilities, which are relevant for particle acceleration, radiation emission and to seed MHD processes at long time-scales.
\end{abstract}



\maketitle

A fundamental question in plasma physics concerns the stability of a given plasma configuration. Unstable plasma configurations are ubiquitous and constitute important dissipation sites via the operation of plasma instabilities, which typically convert plasma kinetic energy into thermal and electric/magnetic field energy. Plasma instabilities can occur at microscopic (particle kinetic) and macroscopic (magnetohydrodynamic, MHD) scales, and are generally studied separately using simplified frameworks that focus on a particular scale and neglect the other. This approach conceals the role that microscopic processes may have on the macroscopic plasma dynamics, which in many scenarios cannot be disregarded. It is now recognized, for instance, that collisionless plasma instabilities operating on the electron scale in unmagnetized plasmas, such as the Weibel \cite{Weibel:1959wg} and streaming instabilities \cite{Bret:2004fh}, play a crucial role in the formation of (macroscopic) collisionless shocks in astrophysical \cite{Silva:2003tq,Frederiksen:2004tl,Nishikawa:2005tk,Spitkovsky:2008wo,Martins:2009vd} and laboratory conditions \cite{Fiuza:2012kd,Stockem:2014fn}. These microscopic instabilities result from the bulk interpenetration between plasmas and are believed to be intimately connected to important questions like particle acceleration and radiation emission in astrophysical scenarios \cite{Medvedev:1999uq,Gruzinov:1999vw}.

Plasma shear flow configurations can host both microscopic and macroscopic instabilities simultaneously, although the former have been largely overlooked. Shear flow settings have been traditionally studied using the MHD framework \cite{Frank:1996wn,Keppens:1999vu,Zhang:2009kb}, where the Kelvin-Helmholtz instability (KHI) is the only instability known to develop \cite{1961hhs..book.....C}. Only very recently have collisionless unmagnetized plasma shear flows been addressed experimentally \cite{Kuramitsu:2012fs} and using particle-in-cell (PIC) simulations, revealing a rich variety of electron-scale processes, such as the electron-scale KHI (ESKHI), dc magnetic field generation, and unstable transverse dynamics \cite{Alves:2012vb,Grismayer:2013bi,Liang:2013gv,Liang:2013ic,Nishikawa:2013bt,Nishikawa:2014wl}. The generated fields and modified particle distributions due to these microscopic processes can strongly impact succeeding macroscopic dynamics of the shear flow. However, whilst both the ESKHI and the dc magnetic field formation mechanism have been treated theoretically, an analytical description of the transverse shear instability, observed in kinetic simulations, is still lacking. An established analytical description is crucial in order to assess its relevance in different physical scenarios.


This Letter focuses on the shear surface instability that occurs in the plane perpendicular to that of the ESKHI. These new unstable modes explain the transverse dynamics and structures observed in PIC simulations in \cite{Alves:2012vb,Liang:2013gv,Liang:2013ic,Nishikawa:2014wl}. We label this effect the mushroom instability (MI) due to the mushroom-like structures that emerge in the electron density.


We analyze the stability of electromagnetic perturbations in the transverse $xy$ plane of a collisionless plasma shear flow with velocity profile $\vec{v_0}=v_0(x)\vec{e}_z$. We assume a cold ($v_{th} \ll v_0$, where $v_{th}$ is the thermal velocity) unmagnetized plasma, and impose charge and current neutrality, $n_{e0}=n_{i0}=n_{0}$, $v_{e0}=v_{i0}$ (subscripts $e$ and $i$ refer to electron and ion quantities, respectively) to guarantee initial equilibrium. We use a two-fluid model where the relativistic equations of motion of the electron and ion fluid are coupled to Maxwell's equations. We assume linear perturbations in the fluid quantities of the form ${f}=f_0+\delta f$ with $\delta f=\bar{\delta f}(x) e^{iky-i\omega t}$ ($\omega\in\mathbb{C}$ and $k\in\mathbb{R}$ are frequency and wavenumber, respectively) with all zeroth order quantities being zero except for $n_0(x)$ and $v_0(x)$. We find, for the perturbed current densities, $\delta j_x =-e(1+m_e/m_i)n_0\delta v_{ex}, \delta j_y =-e(1+m_e/m_i)n_0\delta v_{ey}, \delta j_z =-e(1+m_e/m_i)(n_0\delta v_{ez}+\delta nv_0)$. Substituting in Maxwell's equations, we can simplify the original set of ten coupled equations in a reduced $2\times 2$ system: ${\bf \bar{\bar \epsilon}}.{\bf \delta E}={\bf 0}$, where the coefficient of the tensor $\epsilon_{ij}$ are operators of the form $\epsilon_{ij}=\sum_{m=0}^{2}C_{ij,m}(\omega,k_y,v_0(x),n_0(x))\partial_{x^m}$ and ${\bf \delta E}=(\delta E_y,\delta E_z)$. In order to obtain analytical results we use a step velocity shear and density profile of the form $v_0(x)=v^{-}+(v^{+}-v^{-})\mathcal{H}(x)$ and $n_0(x)=n^{-}+(n^{+}-n^{-})\mathcal{H}(x)$ where $\mathcal{H}$ is the Heaviside function. Integrating ${\bf \bar{\bar \epsilon}}.{\bf \delta E}={\bf 0}$ for $x\neq0$ and using the continuity of $\delta E_y$ and $\delta E_z$ across the shear interface, we find solutions corresponding to evanescent waves: $\delta E_{y,z}(x)=\bar{\delta E}_{y,z}(0)e^{-k_{\perp}^{\pm}|x|}$ where $k_{\perp}^{\pm}=\sqrt{D_{\perp}^{\pm}/c^2}$, $D_{\perp}^{\pm}=c^2k^2+\omega_{pe\pm}^2 \gamma_{\pm}-\omega^2$, $\omega_{pe\pm}^2=e^2n^{\pm}(1+m_e/m_i)/\epsilon_0m_e$ and $\gamma_{\pm}=1/\sqrt{1-(v^{\pm}/c)^2}$. By evaluating the derivative jump of the electric fields across the shear interface we arrive at ${\bf \bar{\bar I}}.{\bf \delta E_0}={\bf 0}$, where ${\bf \delta E_0}=(\bar{\delta E}_y(0),\bar{\delta E}_z(0))$ and $I_{ij}=a_{ij}^{+}k_{\perp}^{+}+a_{ij}^{-}k_{\perp}^{-}$; $a_{11}=(\omega^2-\omega_{pe}^2/\gamma_0)D_{\perp}^{-1}$, $a_{12}=-(kv_0/\omega)(\omega_{pe}^2/\gamma_0)D_{\perp}^{-1}$, $a_{21}=a_{12}$ and $a_{22}=-1-(k^2c^2/\omega^2-1)(\omega_{pe}^2/\gamma_0)(v_0^2/c^2)D_{\perp}^{-1}$. The dispersion relation is finally obtained by solving $\mathrm{det}(\bf \bar{\bar I})=0$. In the special case, $n^{+}=n^{-}=\bar{n}_0$ and $v^{+}=-v^{-}=\bar{v}_0$, the growth rate reads
\begin{equation}
\label{eq:growthrate}
\frac{\Gamma}{\omega_{pe}}=
\frac{1}{\sqrt{2}}
\left(
{\sqrt{\frac{4k^2\bar{v_0}^2}{\bar{\gamma}_0\omega_{pe}^2}+D_{\sslash}^2}-D_{\sslash}}
\right)^{1/2},
\end{equation}
with $\Gamma=\mathrm{Im}(\omega)$ and $D_{\sslash}=1/\bar{\gamma}_0^3+k^2c^2/\omega_{pe}^2$. 

The fastest growing mode of this unstable branch ($\partial_k\Gamma =0$) is found at $k\rightarrow\infty$ (it will be shown later that finite thermal effects and/or smooth velocity shear profiles introduce a cutoff an finite $k$). This limit corresponds to the maximum growth rate of the instability $\Gamma_\mathrm{max}$ for a given shear flow Lorentz factor, and it is given by $\Gamma_\mathrm{max}/\omega_{pe}=\bar{v_0}/c\sqrt{\bar{\gamma}_0}$. In the limit, $n^{-}\gg n^{+}/\gamma_{+}^3$, $\gamma_{+}\gg 1$ and $v^{-}=0$, relevant for astrophysical jet/interstellar medium shear interaction, the maximum growth rate yields $\Gamma_\mathrm{max}/\omega_{pe+}\simeq 1/\sqrt{2\gamma_{+}}$. 

\begin{figure}[t!]
\begin{center}
$$\includegraphics[width=1.0\columnwidth]{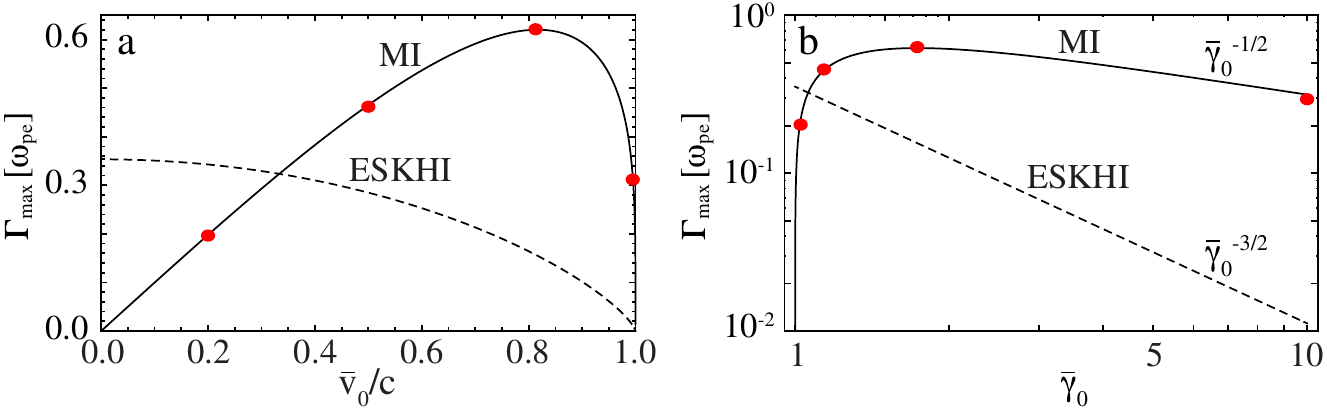}$$ 
\caption{\label{fig:mivskhi} Maximum growth rates of the MI and ESKHI versus (a) flow velocity $\bar{v_0}$ and (b) flow Lorentz factor $\bar{\gamma_0}$, highlighting the dominance of the MI (ESKHI) in relativistic (subrelativistic) flows. The red points represent results of 2D PIC simulations.}
\end{center}
\end{figure}

The MI and ESKHI \cite{Alves:2012vb} growth rates are compared for different shear Lorentz factors and velocities in Figure~\ref{fig:mivskhi}. It is clear that the ESKHI has higher growth rates than the MI for subrelativistic settings. However, the MI growth rate decays with $\bar{\gamma}_0^{-1/2}$, slower than the ESKHI, which decays with $\bar{\gamma}_0^{-3/2}$, as shown in Figure~\ref{fig:mivskhi}.b. Therefore, given that the noise sources for both instabilities are similar, the MI is the dominant electron-scale instability in relativistic shear scenarios.

To verify the theoretical model and better understand the underlying feedback cycle of the MI, we first analyze the evolution of a single unstable mode in an electron-proton ($e^-p^+$) shear flow using the PIC code OSIRIS \cite{Fonseca:2002wg,Fonseca:2008ib}. We simulate a domain with dimensions $20 \times 5  ~ ( c/\omega_{pe} )^2$, resolved with 40 cells per $c/\omega_{pe}$, and use 36 particles per cell per species. The shear flow initial condition is set by the velocity field $\bar{v_0} = +0.5c$ for $L_x/4 < x < 3L_x/4$, and $\bar{v_0} = -0.5c$ for $x < L_x/4 ~\cup~ x>3L_x/4$, where $L_i$ is the size of the simulation box in the $i\mathrm{th}$ direction. The system is initially charge and current neutral. Periodic boundary conditions are imposed in every direction. In order to ensure the growth of a single mode, both $e^-p^+$ temperatures are set to zero, and an initial harmonic perturbation $\delta v_x = \bar{\delta v_x} \mathrm{cos}(k_\mathrm{seed}y)$ in the velocity field of the electrons, with $\bar{\delta v_x} = 10^{-4}c$ and $k_\mathrm{seed}=2\pi/L_y$, is introduced to seed the mode $k_\mathrm{seed}$ of the instability.

\begin{figure}[t!]
\begin{center}
$$\includegraphics[width=1.0\columnwidth]{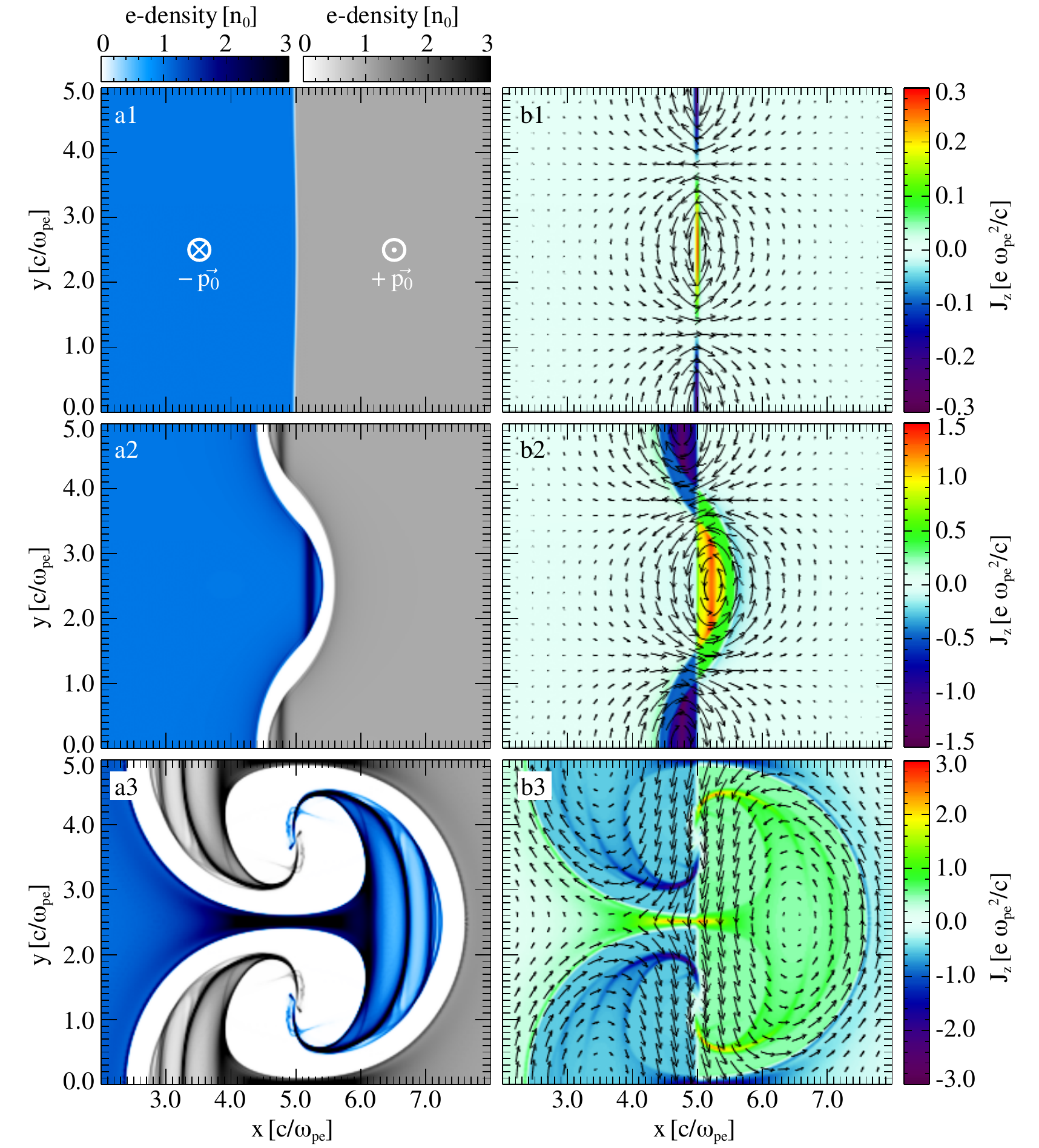}$$ 
\caption{
Evolution of a single unstable mode ($k_\mathrm{seed}=2\pi/5\omega_{pe}/c$) of the MI in an $e^-p^+$ shear. Frames (a) and (b) show the electron density and the out-of-plane current density $J_z$, at times (1) $t = 16/\omega_{pe}$, (2) $t = 26/\omega_{pe}$ and (3) $t = 38/\omega_{pe}$. The evolution of the self-consistent in-plane magnetic field is also displayed in Frames (b).
}
\label{fig:seeded-mush}
\end{center}
\end{figure}

The evolution of the electron density, the out-of-plane current density $J_z$, and the in-plane magnetic field are presented in Figure~\ref{fig:seeded-mush}, which zooms in on the shear interface at $x = L_x/4$. The initial velocity perturbation $\delta v_x$ transports electrons across the velocity shear gradient, producing a current imbalance in $\delta J_z$. This current induces an in-plane magnetic field (namely $\delta B_y$) that, in turn, enhances the velocity perturbation $\delta v_x$ via the $v_0 \times \delta B_y$ force. The enhanced velocity perturbation then leads to further electron transport across the velocity shear gradient in a feedback loop process, which underlies the growth of the instability in the linear stage. The surface wave character of the fields, predicted by the linear theory, is also observed in Figure~\ref{fig:seeded-mush}.b1.

The MI eventually enters the nonlinear phase when the growing magnetic fields become strong enough to significantly displace the electrons and distort the shear interface. The nonlinear distortion of the shear interface in the electron fluid (Figure~\ref{fig:seeded-mush}.a2) leads to the formation of electron surface current filaments (Figure~\ref{fig:seeded-mush}.b2). These surface current filaments effectively translate in a strong dc (non-zero average along the $y$ direction) out-of-plane current structure on either side of the shear interface. Note that the protons in the background remain unperturbed on these time-scales due to their inertia, and also contribute to the formation of the dc current. The dc current structure induces a strong dc magnetic field in $B_y$, as seen by the uniform field lines along the shear interface in Figures~\ref{fig:seeded-mush}.b2-3. The dc magnetic field, in turn, continues to drive the shear boundary distortion via the $v_0 \times \delta B_y$ force, effectively mixing the electrons across the shear, and further enhancing the dc current structure in an unstable loop. The development of the dc magnetic field has been previously shown to be associated to other electron-scale shear flow processes, like the nonlinear development of the cold ESKHI and electron thermal expansion effects \cite{Grismayer:2013bi}. Here we show that the MI is an additional mechanism capable of driving the dc magnetic field. The nonlinear distortion of the shear boundary in the electron fluid, driven by the dc field physics, ultimately gives rise to the formation of the mushroom-like electron density structures shown in Figure~\ref{fig:seeded-mush}.a3. Interestingly, these electron density structures are very similar to those produced by the Rayleigh-Taylor instability \cite{1974PhRvL..33..761B,Mima:1978wd,Takabe:1985ft} despite the great disparity between scales (electron-kinetic versus hydro/MHD scales) and different underlying physics. 

\begin{figure}[t!]
\begin{center}
$$\includegraphics[width=1.0\columnwidth]{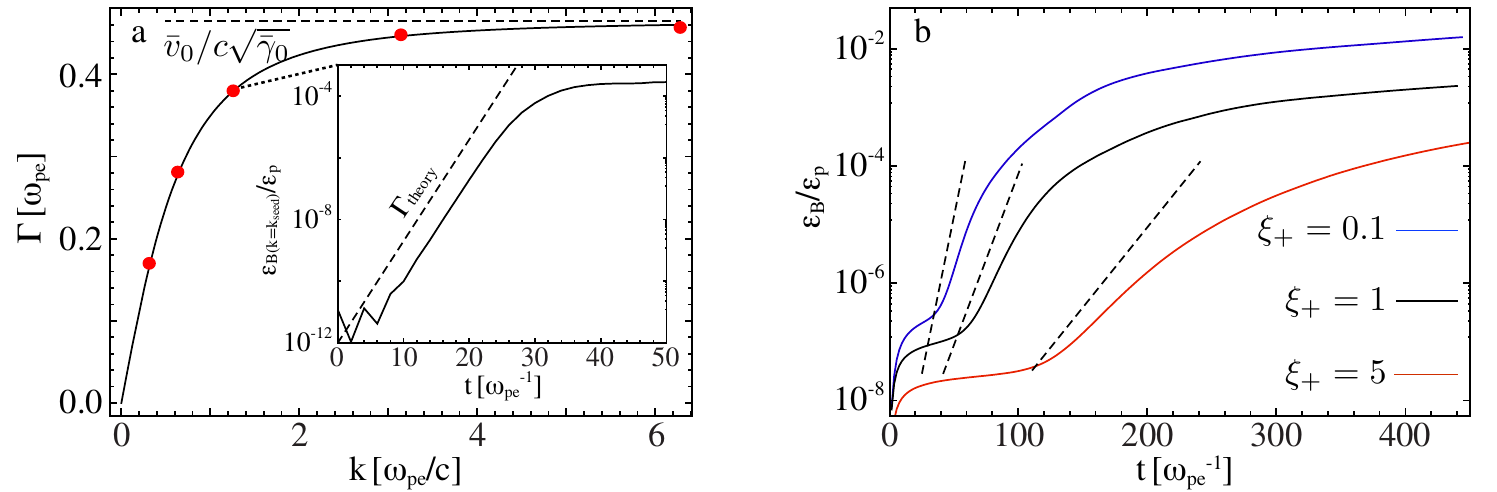}$$ 
\caption{\label{fig:thvssim} a) MI growth rate versus wavenumber for an $e^-p+$ shear flow with $\bar{v^+}=-\bar{v^-}=\bar{v_0}=0.5c$ (solid line); red dots represent results of 2D PIC simulations. The inset shows the evolution of $\epsilon_B/\epsilon_p$ for the single mode simulation with $k_\mathrm{seed}=2\pi/5~\omega_{pe}/c$. b) Finite temperature effects on the growth rate of the MI triggered by a hot relativistic $e^{-}e^{+}$ jet with $\gamma_+=50$, shearing with a cold stationary plasma with $\gamma_-=1$ and $n^+=n^-$. Dashed lines represent the theoretical slope $\Gamma_\mathrm{max/}\omega_{pe+} = 1/\sqrt{2\langle \gamma_+\rangle}$ for each case.}
\end{center}
\end{figure}

The evolution of the ratio of total magnetic field energy to initial particle kinetic energy ($\epsilon_B/\epsilon_p$), for the single mode $k_\mathrm{seed}=2\pi/5~\omega_{pe}c$ simulation, is presented in the inset of Figure ~\ref{fig:thvssim}. The exponential growth associated with the linear development of the instability is observed for $10 \lesssim t\omega_{pe} \lesssim 30$, matching the theoretical growth rate. The instability saturates at $t\omega_{pe} \simeq 40$, approximately when the size of the mushroom-like density structure is on the order of $2\pi/k_\mathrm{seed}$ (Figure~\ref{fig:seeded-mush}.a3). Various simulations with different $k_\mathrm{seed}$ values were performed, verifying the dispersion relation of Eq.~\ref{eq:growthrate} (Figure~\ref{fig:thvssim}.a).

The inclusion of thermal effects should introduce an instability cutoff at a finite wavenumber $k_\mathrm{cutoff}\sim 2\pi/\lambda_D$ ($\lambda_D$ is the Debye length), since the thermal pressure force $F_p \sim k(k_B T_{\perp})$ ($k_B$ and $T_{\perp}$ are the Boltzmann constant and electron temperature perpendicular to $v_0$), can overcome the electric/magnetic forces that drive the unstable arrangement of the electric currents at high $k$ \cite{Silva:2002fz}. Yet such effects must be incorporated through kinetic theory, particularly when $v_{th} \sim \bar{v_0}$, since the thermal expansion of electrons may mitigate the velocity shear gradient, impacting the development of the instability. Interestingly, however, the thermal expansion is greatly reduced when considering ultra-relativistic ($\gamma_0 \gg 1$) shear flows, incidentally the regime where the MI is most significant, allowing the use (to some extent) of the cold fluid treatment outlined above. When considering a plasma of temperature $T_R$ (defined in its rest frame) and drifting relativistically in the $z$ direction with mean velocity $v_0=\beta_0c$, most of the particles have a large Lorentz factor and thus $\langle \beta^2\rangle = \int d^3{\vec p}f({\vec p})\beta^2\simeq 1$, where $f({\vec p})$ is the Juettner distribution \cite{1911AnP...339..856J,WRIGHT:1975uh}. The velocity dispersion in the direction of the drift is $\langle (\beta_z-\beta_0)^2\rangle = \langle \beta_z^2\rangle-\beta_0^2\simeq 0$. Since $\langle \beta^2\rangle=\langle \beta_{\perp}^2+\beta_z^2\rangle$, we find that $\langle \beta_{\perp}^2\rangle \simeq 1-\beta_0^2=1/\gamma_0^2$ (exact for $\xi = k_BT_R/m_ec^2 \gg 1$), restricting the thermal expansion at high $\gamma_0$. The average particle Lorentz factor increases with $T_R$ as $\langle \gamma \rangle \sim \gamma_0(1+(1+\mu(\xi))\xi)$ \cite{WRIGHT:1975uh}, where $\mu(\xi) = 3$ ($3/2$) for $\xi \gg 1$ ($\xi \ll 1$); this factor modifies the growth rate of the MI to $\Gamma_\mathrm{max}/\omega_{pe}\sim 1/\sqrt{\langle \gamma \rangle}$ (for the symmetric shear scenario) due to the enhanced average relativistic particle mass. Hence, thermal expansion effects remain negligible as long as $c\sqrt{\langle \beta_\perp^2\rangle}/\Gamma_\mathrm{max} \ll c/\omega_{pe}\sqrt{\langle \gamma \rangle}$ \cite{Grismayer:2013bi}, which implies $\gamma_0\gg 1$.

\begin{figure}[t!]
\begin{center}
$$\includegraphics[width=1.0\columnwidth]{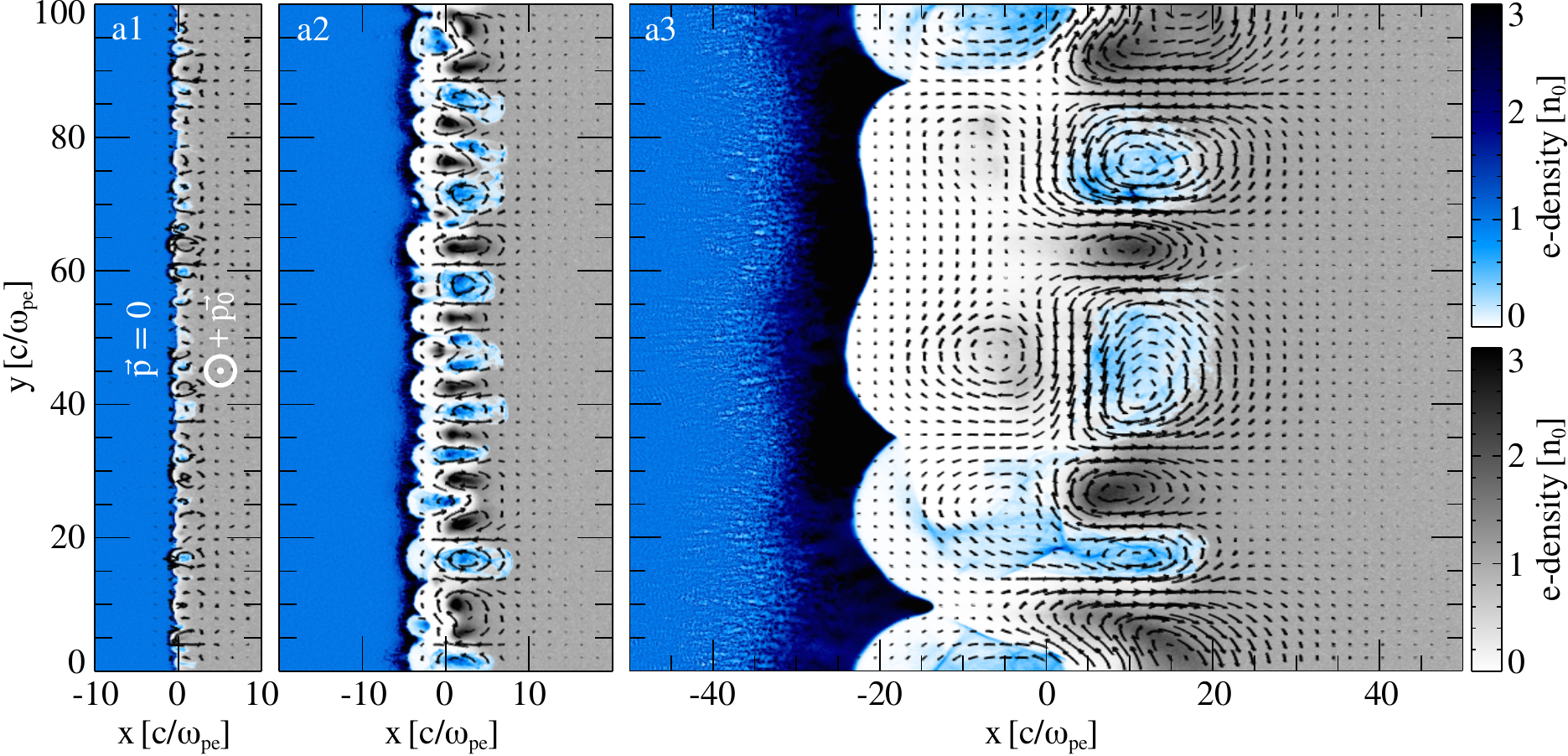}$$ 
\caption{
Electron density evolution in relativistic $e^-e^+$ shear, with $\gamma_+=50$ and $\gamma_-=1$, at (a1) $t=80/\omega_{pe}$, (a2) $t=160/\omega_{pe}$ and (a3) $t=400/\omega_{pe}$. Small-scale current filaments are excited at early times, merging into large-scale structures as the instability develops. Vector field represents self-generated magnetic field structure.}
\label{fig:thermal-mush}
\end{center}
\end{figure}

We have verified the role of thermal effects on the MI development in an $e^-e^+$ shear configuration consisting of a hot relativistic jet, with $\gamma_+ = 50$ and $\xi_+=0.1, 1, 5$, and a cold stationary background, with $\gamma_- = 1$ and $\xi_- = 10^{-4}$; we consider $n^-=n^+=n_0$. We simulate these configurations in a domain of $200 \times 100 ~(c/\omega_{pe})^2$ resolved with a $4000\times2000$ grid. As expected, the MI growth rate decreases with $\xi_+$ (Figure~\ref{fig:thvssim}.b), and is consistent with the temperature-enhanced average relativistic particle mass predicted analytically. The evolution of the electron density around one of the shear interfaces of the system is presented in Figure~\ref{fig:thermal-mush} for the case $\xi_+=1$. At early times, $k\lesssim k_{\textrm{cutoff}}$ structures are observed in the electron density in Figure~\ref{fig:thermal-mush}.a1. These small-scale structures, which are essentially surface current filaments, magnetically interact with each other and merge (Figure~\ref{fig:thermal-mush}.a2-3), forming larger-scale structures in a similar manner to the current filament merging dynamics associated with the current filamentation instability \cite{2005ApJ...618L..75M}. The evolution of the self-generated magnetic field structure is also illustrated in Figure~\ref{fig:thermal-mush}. In addition, saturation of the self-generated magnetic field ($B_\mathrm{sat}$) is achieved when $\Gamma_\mathrm{max}\sim\omega_B=e B_\mathrm{sat}/m \langle \gamma_+ \rangle$, based on magnetic trapping considerations \cite{Davidson:1972jg}.


We have also considered the effects of a smooth shear in the development of the MI. We have performed 2D PIC simulations with velocity profiles of the form $v_0(x)= \bar{v_0}~\mathrm{tanh}(x/L_v)$, where $L_v$ is the shear gradient length, which reveal the persistent onset of the MI at gradient lengths up to $L_v = c/\omega_{pi} \gg c/\omega_{pe}$ (where $\omega_{pi}$ is the ion plasma frequency). The MI can thus be of relevance to the physics on the ion length/time scale. It is found that the introduction of the finite shear length $L_v$ also leads to a fastest growing mode at a finite $k_\mathrm{max} \simeq 2\pi/L_v$. The growth rate $\Gamma_\mathrm{max}$ of the MI decreases with increasing $L_v$, but remains higher than the ESKHI for the same $L_v$ in relativistic scenarios.

\begin{figure}[t!]
\begin{center}
$$\includegraphics[width=1.0\columnwidth]{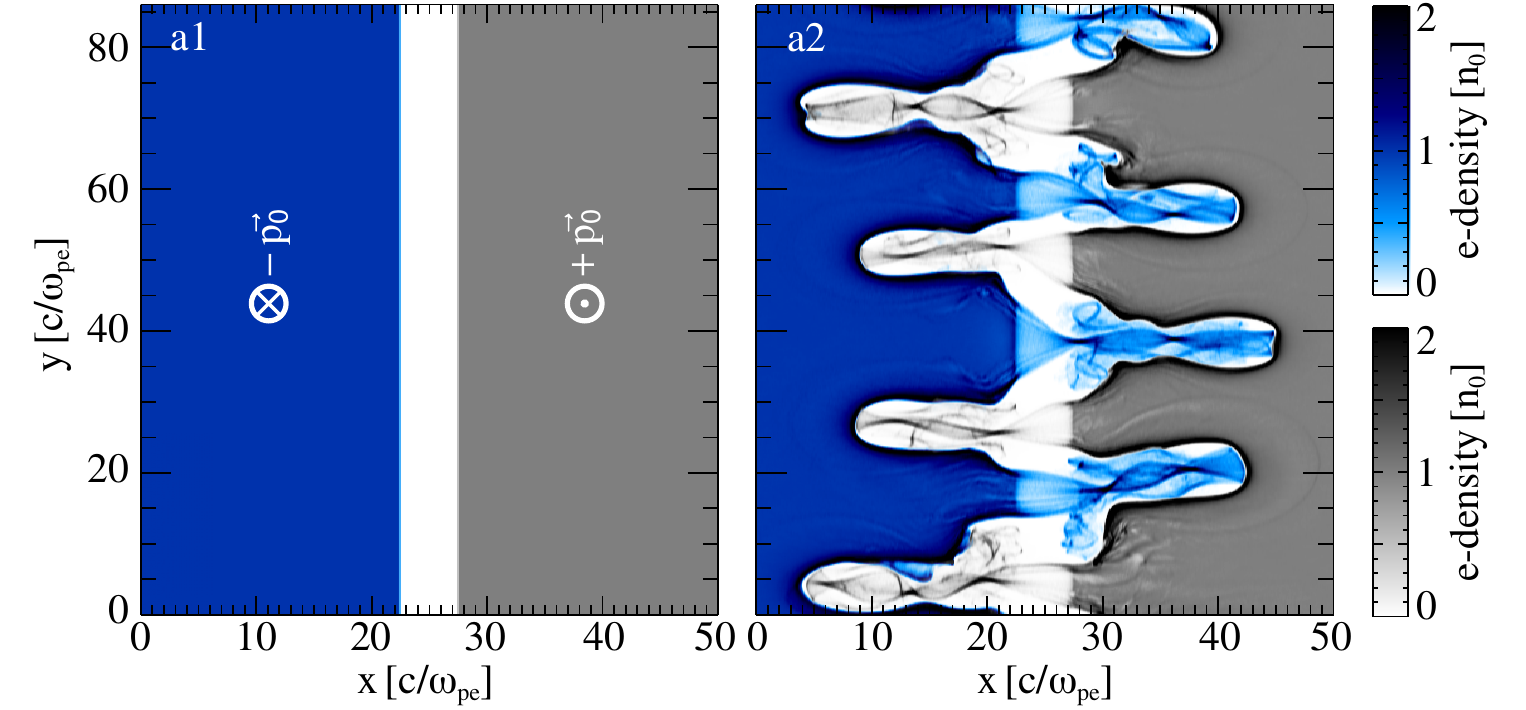}$$ 
\caption{
MI development in the case of a finite gap between flows with $L_\mathrm{gap}=5~c/\omega_{pe}$ and $\bar{v_0}=\sqrt{2/3}~c$. Frames a) and b) reveal the electron density at times $t=0/\omega_{pe}$ and $t=725/\omega_{pe}$, respectively.
}
\label{fig:mushgap}
\end{center}
\end{figure}

Interestingly, due to its electromagnetic nature, the MI is found to operate in the absence of contact between flows, i.e., in the case of a finite gap $L_\mathrm{g}$ separating the shearing flows (Figure~\ref{fig:mushgap}), highlighting the different nature of the MI compared to the (bulk) two-stream and current filamentation instabilities. This setting is closely connected to the work explored in \cite{Silveirinha:2014bd} on the development of optical instabilities in (subrelativistic) nanoplasmonic scenarios, consisting of shearing metallic slabs separated by a nanometer-scale gap; the development of such instabilities results in an effective non-contact friction force between slabs \cite{1997JPCM....910301P,Silveirinha:2014ec}. The role of the MI, however, was overlooked since only subrelativistic configurations were considered; the transverse MI mode will be predominant in the relativistic regime. The new MI modes are found by taking into account the new boundary conditions imposed by the gap. The new eigenmodes of the system have a more complex spatial structure; the surface waves peak at the flow boundaries, evanesce with $k_\perp\mid_{n_0 = n_{e0}}$ and $k_\perp\mid_{n_0 = 0}$ in the plasma and vacuum regions, respectively, and couple in the gap. Therefore, the interaction between flows across the gap is strong as long as $L_\mathrm{g} k_\perp\mid_{n_0 = 0} \sim 1$. Simulations show that the most unstable mode is found at $k_\mathrm{max} \sim 2\pi/L_g$ and that the instability growth rate decreases as the gap is broadened, $\Gamma_\mathrm{max}^\mathrm{gap} \sim \Gamma_\mathrm{max}~\mathrm{exp}(-L_g\omega_{pe}/c\sqrt{2})$.


In conclusion, we have described a fast-growing electron-scale instability (MI) that develops in unmagnetized shear flows, and that explains the structures observed in \cite{Alves:2012vb,Liang:2013gv,Liang:2013ic,Nishikawa:2014wl}. We have shown via analytic theory that the MI grows faster than the closely related ESKHI (that operates in the perpendicular plane) in relativistic shears. These analytical results support the findings reported in \cite{Alves:2012vb}, where 3D PIC simulations of cold, unmagnetized shear flows were investigated, showing the simultaneous operation of the ESKHI and MI, and where it was found that the MI dominates over the ESKHI in relativistic scenarios. Furthermore, we have presented 2D PIC simulations that give insight into the formation of the mushroom-like electron density structures in the nonlinear stage of the MI. 
Finally, the MI was shown to be robust against a variety of non ideal (realistic) shear conditions: (i) operating under relativistic thermal effects, (ii) different densities between shearing flows, (iii) smooth velocity shear profiles ($L_v \gg c/\omega_{pe}$), and, quite surprisingly, (iv) even in the absence of contact between flows. Relativistic shear flow conditions may be reproduced in the laboratory by propagating a globally neutral relativistic $e^-e^+$ beam \cite{Muggli:2013wh} in a hollow plasma channel \cite{Schroeder:1999wy,Yi:2014fh}, allowing experimental access to the MI and shear flow dynamics on the electron scale. This will be explored and presented elsewhere.
In the astrophysical context, the unexpected ability of these microscopic shear instabilities (MI and ESKHI) to generate strong (equipartition) macroscopic ($\gg c/\omega_{pe}$, in the case of $e^-p^+$ shears) fields from initially unmagnetised conditions can directly impact particle acceleration and radiation emission processes, and can seed the operation of macroscopic (MHD) processes at later times (e.g. magnetic dynamo).

E. P. Alves and T. Grismayer contributed equally to this work. This work was partially supported by the European Research Council ($\mathrm{ERC-2010-AdG}$ Grant 267841) and FCT (Portugal) grants SFRH/BD/75558/2010 and IF/01780/2013. We acknowledge PRACE for awarding access to SuperMUC based in Germany at Leibniz research center. Simulations were performed at the IST cluster (Lisbon, Portugal), and SuperMUC (Germany).


\bibliographystyle{apsrev4-1.bst}
\bibliography{mypapers_16_4_2015.bib}


\newpage

\end{document}